\documentclass{aastex}

\def\arcmin{\hbox{$^\prime$}}
\def\arcsec{\hbox{$^{\prime\prime}$}}

\def\farcm{\hbox{$.\mkern-4mu^\prime$}}
\def\farcs{\hbox{$.\!\!^{\prime\prime}$}}

\usepackage{graphicx}
\usepackage{rotating}
\usepackage{longtable}

\shorttitle{Proper motions in 30 Doradus}
\shortauthors{PLATAIS ET AL.}

\begin{document}

\title
{{\it HST} astrometry in the 30 Doradus region: II. Runaway stars from
new proper motions in the Large Magellanic Cloud}
\author{Imants Platais\altaffilmark{1},
Daniel J. Lennon\altaffilmark{2},
Roeland P. van der Marel\altaffilmark{3,1},
Andrea Bellini\altaffilmark{3}, 
Elena Sabbi\altaffilmark{3},
Laura L. Watkins\altaffilmark{3},
Sangmo T. Sohn\altaffilmark{3},
Nolan R. Walborn\altaffilmark{3},
Luigi R. Bedin\altaffilmark{4},
Christopher J. Evans\altaffilmark{5},
Selma E. de Mink\altaffilmark{6},
Hugues Sana\altaffilmark{7},
Artemio Herrero\altaffilmark{8},
Norbert Langer\altaffilmark{9},
Paul Crowther\altaffilmark{10}
}
\email{imants@jhu.edu}
\altaffiltext{1}{Department of Physics and Astronomy, Johns Hopkins University, 3400 North Charles Street, Baltimore, MD 21218, USA}
\altaffiltext{2}{European Space Astronomy Centre, Camino bajo del Castillo, Urbanizacion Villafranca del Castillo, Villanueva de la Ca\~{n}ada, 28692, Madrid, Spain}
\altaffiltext{3}{Space Telescope Science Institute, 3700 San Martin Drive, Baltimore, MD 21218, USA}
\altaffiltext{4}{INAF-Osservatorio Astronomico di Padova, Vicolo dell'Osservatorio 5, I-35122 Padova, Italy}
\altaffiltext{5}{UK Astronomy Technology Centre, Royal Observatory Edinburgh,
Blackford Hill, Edinburgh, EH9 3HJ, UK}
\altaffiltext{6}{Anton Pannekoek Institute for Astronomy, University of Amsterdam, Science Park 904, 1098 XH, Amsterdam, The Netherlands}
\altaffiltext{7}{Institute of astrophysics, KU Leuven, Celestijnlaan 200D, 3001, Leuven, Belgium}
\altaffiltext{8}{Instituto de Astrof\'{i}sica de Canarias, 38200 La Laguna, Tenerife, Spain}
\altaffiltext{9}{Argelander-Instit\"ut f\"ur Astronomie, Universit\"at Bonn, Auf dem H\"ugel 71, 53121, Bonn, Germany}
\altaffiltext{10}{Department of Physics and Astronomy, Hicks Building, Hounsfield Road, University of Sheffield, Sheffield S3 7RH, United Kindom}

\begin{abstract}
We present a catalog of relative proper motions for 368,787 stars in
the 30 Doradus region of the Large Magellanic Cloud (LMC), based
on a dedicated two-epoch survey with the  Hubble Space Telescope
({\it HST}) and supplemented with proper motions from our pilot archival study.
We demonstrate that a relatively short epoch difference of 3 years
is sufficient to reach a $\sim$0.1 mas~yr$^{-1}$ level of precision or better.
A number of stars have relative proper motions exceeding
a 3-sigma error threshold, representing a mixture of Milky Way
denizens and 17 potential LMC runaway stars. Based upon 183 VFTS
OB-stars with the best proper motions, we conclude that none of them move
faster than $\sim$0.3 mas~yr$^{-1}$ in each coordinate --
 equivalent to $\sim$70 km~s$^{-1}$.
Among the remaining 351 VFTS stars with less accurate proper motions,
only one candidate OB runaway can be identified. We rule out any OB star
in our sample moving at a tangential velocity exceeding $\sim$120 km~s$^{-1}$.
The most significant result of this study is finding 10 stars over wide
range of masses, which
appear to be ejected from the massive star cluster R\,136 in the tangential
plane to angular distances from $35\arcsec$ out to $407\arcsec$, equivalent
to 8-98 pc. The tangential velocities of these runaways appear to be
correlated with apparent magnitude, indicating a possible dependence
on the stellar mass.

\end{abstract}


\section{INTRODUCTION}
\label{intro}

This study addresses new measurements of proper motions with {\it HST}
for individual stars in the 30~Doradus (hereafter 30~Dor) area of the LMC
in the context of a search for runaway OB stars and potential scenarios of
how the massive OB stars formed.
In our first paper by \citet{pl15}, hereafter Paper~I, we give the
motivation to study this area and provide a detailed account how to obtain
high-accuracy positions and relative proper motions with various
{\it HST} imaging instruments. We demonstrate that it is possible to measure
reliably an individual relative proper motion down to $\sim$0.1 mas~yr$^{-1}$,
which corresponds to $\sim$25 km~s$^{-1}$ at the distance of the LMC.
As a result, we presented a pilot catalog of positions and relative
proper motions, derived from a targeted single-epoch survey combined with
numerous archival {\it HST} observations spanning up to 17 years. Although 
the precision of proper motions can be as good as $\sim$20~$\mu$as~yr$^{-1}$,
the accuracy of proper motions appears to be significantly lower
due to residual systematic errors.  In addition, as indicated by Fig.~13 of
Paper~I, our pilot catalog has spatial discontinuities and covers only
$\sim$30\% of the available contiguous area.

In Paper~I, we also attempted to identify possible runaway OB stars using
the calculated proper motions for 86,590 stars. That resulted in six
candidate OB proper-motion runaway stars. Interestingly, three of them are part of the VLT-FLAMES Tarantula Survey \citep[VFTS;][]{ev11}, whereas, another
three are additional photometric OB stars. We noted that star VFTS 285
appears to have its proper motion and position consistent with the ejection
scenario from the massive star cluster R\,136. Still, none of the astrometric
candidate OB runaway stars could be considered as a conclusive case.

There are two studies of line-of-sight (LOS) velocities that also address
potential OB runaway stars in the 30 Dor area and slightly eastward of
it \citep{ev15,eva15}. These authors identify a total of 18 candidate
runaway stars. Assuming that none of them is a large-amplitude binary,
only five of them have an excess LOS velocity in the range
$75 <$ v$_{\rm LOS}$ $< 108$ km~s$^{-1}$, while the rest have lower
velocities with the lowest one at 40~km~s$^{-1}$. This  range of excess
LOS velocities implies that the expected total proper motion of true OB runaway
stars may not exceed $\sim$0.4 mas~yr$^{-1}$. The anticipated upper limit is
still non-trivial to measure with {\it HST} over a time span of a few years.
The concept of effective Point-Spread-Function (ePSF),
accurate accounting for geometric distortions, and empirical correction for the effect of Charge Transfer Efficiency (CTE) losses are the three crucial
developments that now allow us to measure relative positions of stars
to the level of $\sim$0.5 mas \citep{an00,an03,an10b}.

In addition, we designed the second epoch observations
such that they matched the first-epoch observations as closely as possible, thus
minimizing the contribution by the main source of systematic errors related
to a star's location on the detector and possible changes in signal-to-noise
ratio between the epochs. The combined set of first- and second-epoch
observations is analyzed here. There is an  overlap in terms of
the applied techniques and methods of analysis with Paper~I and the reader
is frequently directed to this paper. This is also true for the Introduction --
a more detailed scientific motivation is provided in Paper~I. 

\section{{\it HST} SURVEY OF 30 DOR AND DATA REDUCTIONS}\label{hstobs}

One of the main goals in our astrometric survey was to use the Wide Field
Camera 3 (WFC3/UVIS) and the Advanced Camera for Surveys (ACS/WFC) in
parallel in order to cover as much of 30 Dor as possible, focusing in
particular on the VFTS stars \citep{ev11}.
The first epoch
observations (GO-12499; PI: D.~Lennon) were mainly obtained in 2011
October 3-8, while nearly indentical second epoch observations (GO-13359)
were repeated three years later in 2014, October 6-11. Details of the first
epoch observations are given in Paper~I.  For the second-epoch observations, the
last sub-pointing $D$ in each observational set (see Fig.~3, Paper~I) has
a small $\sim$$3\%$ adjustment to its exposure time and some pointings have
different guide stars. As opposed to the first-epoch observations, in 2014
we did not apply preflash to ACS short exposures.
For all practical reasons, both first and second epoch observations
are nearly identical. Altogether, there is a total of 149 ACS/WFC and the same
number of WFC3/UVIS frames, both sets obtained through filter F775W.
Similar to Paper~I, we used \texttt{\_flc.fits} files. The latter are
corrected for the effect of charge-transfer efficiency (CTE) losses in
images \citep[][]{an10b}. 
We note that these corrections reflect the status
of adopted pipeline procedures in the year 2015. At the time of this writing,
some of them have been updated, thus now resulting in slightly different
output of \texttt{\_flc.fits} files.

It should be mentioned that the Hubble Tarantula Treasury Project
\citep[HTTP,][]{sa16} is a rich source of additional observations
in the 30~Dor area. Although the instrumental setup is identical
to our program and include other filters such as F555W and F658N,
there are differences in the visit pointings and orientation angles. This
may introduce unwanted systematics if compared to our observations,
which are {\it optimized} to exclude systematics in proper motions.
Since these HTTP observations do not extend the available span of
time coverage, we decided not to use them in our analysis.

The object detection, the calculation of their centroid, flux, quality
parameter \texttt{qfit}, and correction for geometric distortion -- all
followed the guidelines provided in Paper~I, Sect. 3.1. 

\subsection{Differential charge-transfer inefficiency}\label{diffcti}

Complementary to the CTE losses is the charge-transfer inefficiency
(CTI) effect which is just another way of interpreting small offsets
in the positions introduced during the CCD readout process. Since all
frames have been already corrected for the effect of CTE losses, we
would expect that the residuals from a transformation of the second-epoch
positions into the first-epoch positions do not contain any dependence
on the magnitude of stars in the direction of CCD readout. If this is not true,
then we name such a dependence to be a differential CTI, which in essence
reflects the degree of CTE-correction efficiency.

First, we examined pairs of identical WFC3/UVIS first-to-second epoch pointings
taken with the same exposure time. It is expected that a linear
transformation (offset, rotation, and scale) of distortion-uncorrected
centroids from one epoch to another would enable us to characterize
the stability of star pixel coordinates over time. Due to the large
distance of
the LMC, the expected proper motions over three years have a very limited
contribution to the total budget of positional errors. This
exercise brought to light  two issues: a small offset (a few hundredths of
a pixel) between the two CCD chips and a slope in proper motions as
a function of magnitude and the $Y$ pixel location. The same pattern was
discovered in the similar pairs of ACS/WFC frames. Small variations
in the gap size between the chips are expected but they have no impact on
proper motions because, in our final adjustement, each chip is transformed
into the astrometric reference frame separately (Paper~I, Sect. 3.2).
The second issue is more serious. A slope in the $Y$ pixel location as
a function of magnitude appears to be a hallmark of some residual
(differential)  unaccounted-for
CTE losses. It may appear odd to have this effect after the pipeline
corrections for the CTE losses. However, \citet{an10b} caution that the adopted
empirical model for corrections may not be perfect. By comparing positions
at, e.g., two epochs, we can actually test whether the adopted model
might need a time-dependent component. 

In order to build a statistically significant sample, we selected three
sets of frames -- 15 pairs taken with short exposure (35~s for WFC3/UVIS)
of sub-pointing A (see Paper~I, Fig.~3), 15 pairs with long exposures
(699~s for WFC3/UVIS) of
the same sub-pointing, and 14 pairs of sub-pointing D with somewhat
shorter exposure times (490-507~s for WF3/UVIS). Similarly, we selected the
corresponding ACS/WFC pointings and exposures (see Paper~I, Sect. 2).
Then, for each pair we used least-squares minimization and a linear
three-term polynomial to transform first-epoch positions into the
system of second-epoch positions. In this transformation only positions
of optimally-exposed stars were used (instrumental magnitudes in the
range $-14 < m_{\rm F775W} <-8$) and obvious outliers ignored.
In the next step, residuals of all matching stars were collected and
assigned to the applicable CCD chip, which include precise pixel-location
on that chip and measured instrumental magnitude.

\begin{figure}
{\includegraphics[scale=0.75]{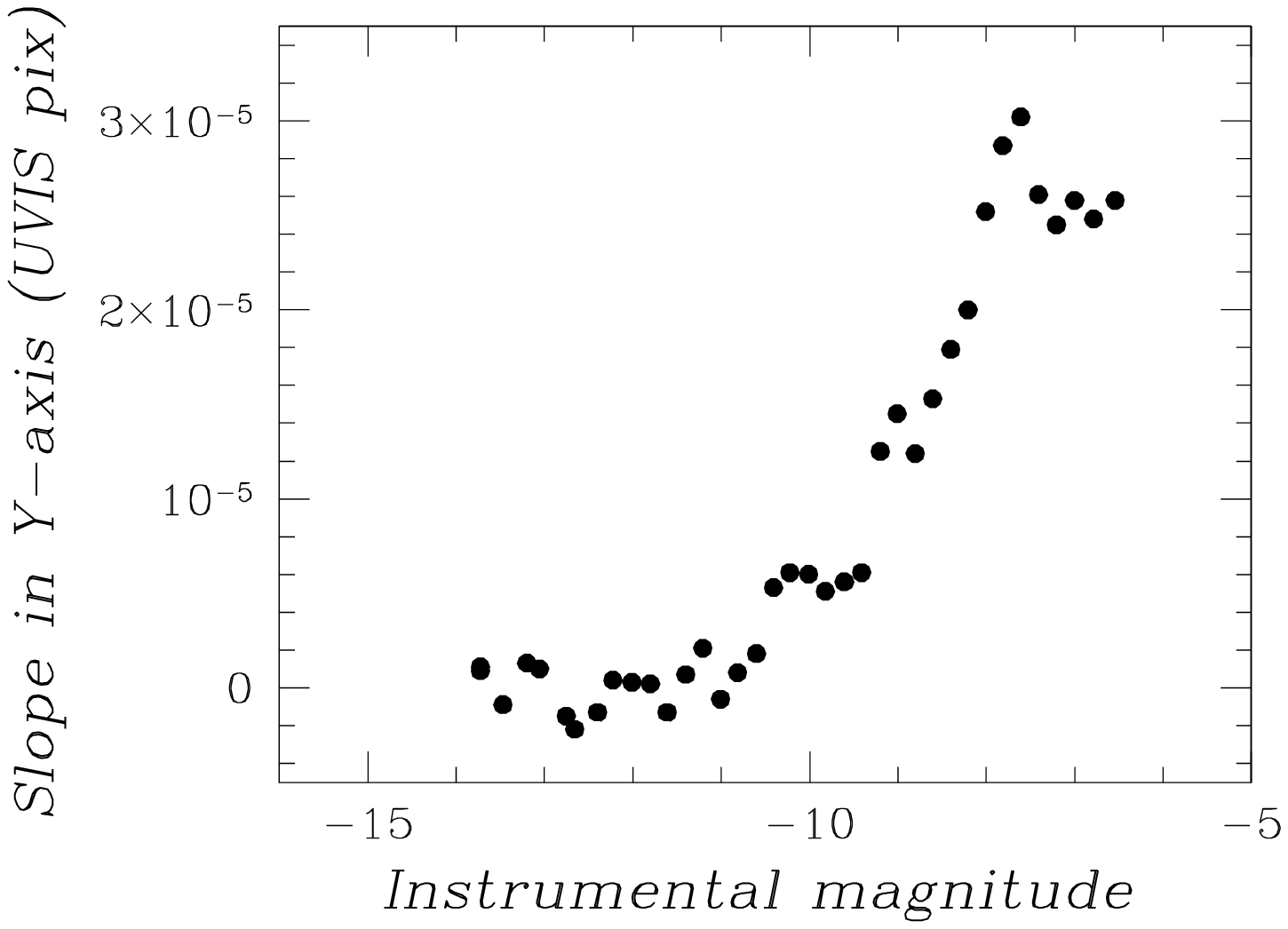}}
\caption{Example of differential-CTE slope distribution as a function of
instrumental magnitude. It is for WFC3/UVIS Chip~1 and short exposures
only, which produced a total of nearly 20K residuals.
}
\label{fig:exa_cte}
\end{figure}

To characterize differential CTI between the two epochs, the collection of
individual residuals for each chip was re-distributed into successive
bins of instrumental
magnitude. The total number of available residuals per chip varies between
$\sim$15,000 and $\sim$150,000 depending on the imager (WFC3 or ACS)
 and the length of exposure time.
Each bin is at least 0.333 mag wide and is shifted by 0.2 mag
with respect to the adjoining bin. The minimum number of residuals per bin
is adopted to be 105. If a bin contains fewer residuals, it is enlarged
until the required number is reached. This procedure is essential for
the bright end of magnitude bins, where the number of stars is always low.
Then, in each bin a least-squares fit is applied to the residuals
in $Y$-coordinate as a function of $Y$. We used a linear 3-term polynomial
which provides a potential slope in these residuals characterizing
differential CT. We could not find statistically significant non-linearity
in the actual fits. An example provided in Fig.~\ref{fig:exa_cte} shows
the general pattern of slopes as a function of magnitude, which reflects
the presence of differential CTI. If this is ignored, then it will
introduce a bias in the calculated proper motions.

\begin{figure}
{\includegraphics[scale=0.75]{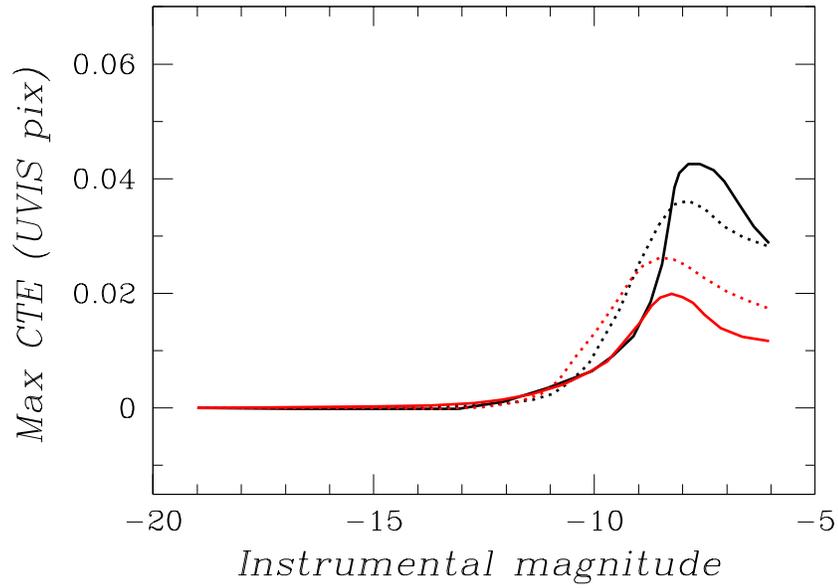}}
\caption{Maximum effect of differential CTE for ACS/WFC as a function of
magnitude, chip selection, and exposure length. Solid lines represent
WFC1; dotted lines -- WFC2. Black lines shows the effect for
short 32~s exposures; red lines -- for long 640~s exposures.
}
\label{fig:acs_cte}
\end{figure}

\begin{figure}
{\includegraphics[scale=0.75]{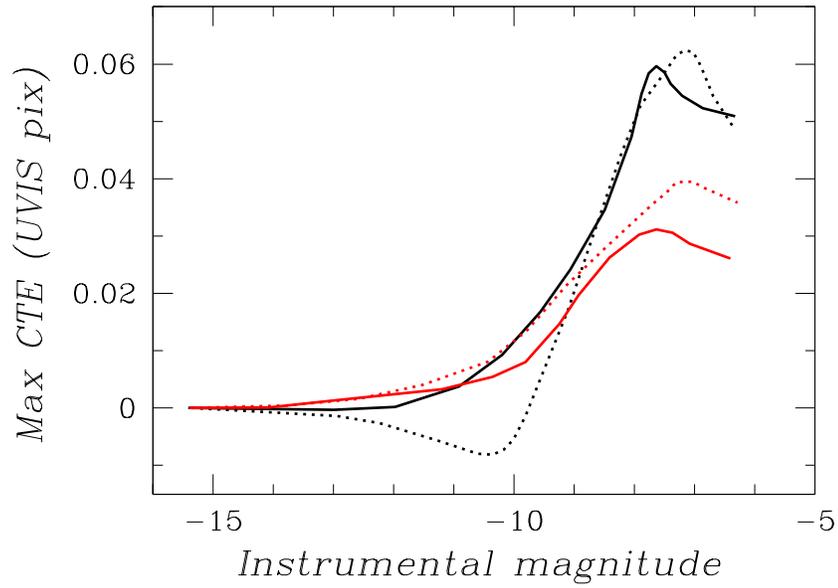}}
\caption{Maximum effect of differential CTE for WFC3/UVIS as a function of
magnitude, chip selection, and exposure length. Solid lines represent
Chip~1; dotted lines -- Chip~2. Black lines shows the effect for
short 35~s exposures; red lines -- for long 699~s exposures.
}
\label{fig:wfc3_cte}
\end{figure}

In Figures~\ref{fig:acs_cte},\ref{fig:wfc3_cte} we show the maximum effect
(at the far edge of a CCD chip) of differential CTI for both imaging
instruments for
short and long exposures. Formally, the effect is dependent on the exposure
time and is more pronounced on short exposures. The turnover
at instrumental magnitude $m_{\rm F775W}~\sim~-8$  appears to be an artifact
of the pixel-based pipeline correction applied to our images.  We hope
that these findings will provide stimulus to the further improvements in
minimization of CTI effects on fluxes and positions. For this project,
we applied differential CTI corrections to all second epoch frames.

In order to apply these corrections to distortion-uncorrected 
coordinates, we generated a total of 12 empirical CTI curves.
That is, each one for a short, long, and an intermediate length
(364-507~s) exposure; for each imaging instrument (ACS/WCS and WFC3/UVIS);
and for each chip. We note that the shape and amplitude of the corresponding
curves for long and intermediate exposures are quite similar, indicating that
the differential  CTI effect may not be linearly-correlated with
the length of exposure.  
The raw distribution of slopes as a function of magnitude
(Fig.~\ref{fig:exa_cte}) is too noisy and coarse to work with. Therefore,
we iteratively smoothed these distributions and then applied cubic
splines to parameterize the resulting curves (see Figs.~\ref{fig:acs_cte},
\ref{fig:wfc3_cte}). For intrumental magnitudes
of $-15$ and brighter a zero correction was adopted. Also, differential CTI
correction of any magnitude is adopted to be zero at the near-edge
of a CCD readout
direction. Once positions of the second-epoch frames were corrected
for differential CTI, the resulting coordinates were then corrected for
geometric distortion using the same routines and parameters as in Paper~I.

We also explored the possibility of some differential CTI effect in the
serial direction because \citet{an10b} reported the presence of the
so-called $x$-hook in the direction away from the readout amplifier.
None was detected, although we did not attempt to explore that for
each separate CCD amplifier.

\subsection{New Proper Motions}\label{promo}

\begin{figure}
{\includegraphics[scale=0.75]{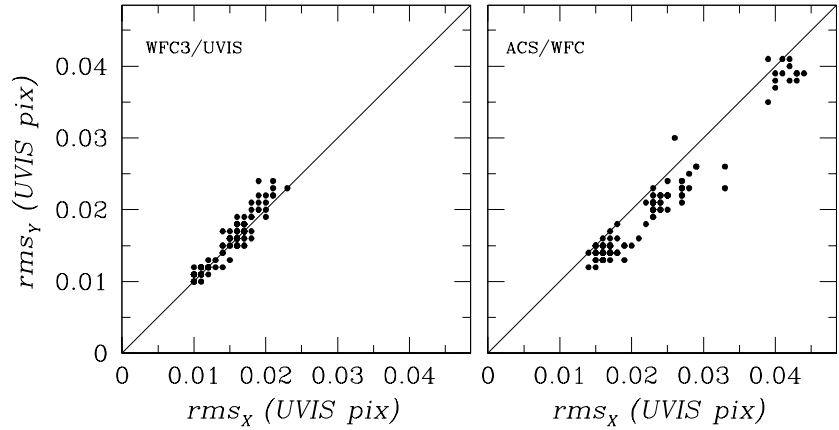}}
\caption{Distribution of rms from the least-squares linear transformation of
individual frames into the astrometric reference catalog. The reason of
the worse rms for ACS/WFC frames is discussed in Sect.~\ref{promo}. There
is a total of 149 rms estimates for each of the {\it HST} imaging instruments.
}
\label{fig:rms_all}
\end{figure}

In Paper~I we described all basic steps on how to calculate proper motions using
mosaic-like observations with three different {\it HST} imaging instruments.
A central role in these calculations served the astrometric reference catalog,
which covers the entire field-of-view (FOV). At the time of constructing
this reference frame, we noticed a need for nonlinear terms in the
transformation of the ACS/WFC positional catalog into the WFC3/UVIS-based
positions and some semi-periodic systematics after this transformation.
Therefore, with the arrival of second-epoch {\it HST} observations in 2014, we
decided to use the WFC3/UVIS positional catalog as is, not including
positions from the slightly overlapping parts of the ACS/WFC reference
catalog. As a result,
all WFC3/UVIS frame-solutions into its own reference
catalog have significantly better rms for overlapping visits (pointings)
01-07 (see Fig.~2, Paper~I). This is confirmed by the distribution of
corresponding rms in both axes (Fig.~\ref{fig:rms_all}). All WFC3/UVIS solutions
yield small rms estimates and never exceed 0.025 pixel. We note that
the second-epoch observations also include the contribution of proper motions
over 3 years. Therefore, the second epochs clump near the rms of $\sim$0.02
pixel. 

Similar solutions for the set of ACS/WFC frames show a noticeably different
pattern. The rms of the first-epoch solutions, as expected, clump near 0.015
pixel but the second-epoch ACS/WFC frames obtained with 32~s exposure cluster
around the considerably higher $\sim$0.04 UVIS pixel.
Clearly, this is not due to the
proper motions.  First, we noticed that the image quality parameter
\texttt{qfit} for nearly all ACS second-epoch frames is significantly
elevated -- up to twice that of the first-epoch relatively bright
images. Despite using the same fiducial ePSFs as in
Paper~I, somehow the actual second-epoch stellar images may have slightly
changed their shape, especially on frames with exposure time of 32~s,
which are least appropriate to construct a new set of ePSFs. Second,
the available correction for geometric distortion of ACS/WFC frames may
not be optimal as indicated by a necessity to apply a quadratic term
in order to align with the WFC3/UVIS positions in constructing
the astrometric reference catalog (Paper~I). As a result, proper motions
from ACS/WFC observations have somewhat lower precision and
accuracy than those derived from the WFC3/UVIS observations.
To alleviate the impact of potential residual geometric distortion,
pixel positions on each chip for both, WFC3/UVIS and ACS/WFC, were
separately translated into  the astrometric reference frame.  

New proper motions were calculated following the instructions given
in Paper~I, Sect. 3.3.2. As in Paper~I, we used the same,
effectively, a four-pixel box to find all common positions for a star on
the system of the astrometric reference catalog. Since the epoch difference
is three years, the chosen box size misses fast moving stars
exceeding $\sim$25 mas~yr$^{-1}$ in one coordinate. Likewise, we have
proper motions from linear least-squares fits to the unweighted and weighted
$XY$ positions. In the final catalogs of proper motions only the ''weighted"
version of the proper motions is provided. 
The principal improvement in our current version of the proper motions
is a complete coverage of the entire available FOV, while the spatial
coverage in Paper~I is patchy and represents only $\sim$30\% of the total FOV.
 
The mean error of proper motions
from WFC3/UVIS observations and for magnitude range $14 < m_{\rm F775W} < 20$
is 0.088 mas~yr$^{-1}$ in either coordinate. This magnitude range
contains all optimally-exposed OB stars.
Similarly, for ACS/WFC observations in the same magnitude range, the mean
error is 0.12 and 0.10 mas~yr$^{-1}$ along RA and Dec, respectively.
These larger errors as well as possible systematics in our astrometry 
lower the chances to detect OB runaways using the ACS/WFC data.

\subsection{Catalogs of proper motions}\label{catalogs}

Measuring {\it individual} proper motions in the LMC is close to the limits
of any ground- or space-based facility's state-of-the-art status,
including the {\it Gaia} mission. Therefore, here we provide four catalogs
of proper motions that give an opportunity to examine their internal
consistency and other properties (Proper Motion Catalog, electronic Table).
All our proper motion measurements are relative and effectively local.
The mean motion
of stars from the astrometric reference catalog should be zero. These
proper motions might be used to establish upper limits on the internal
velocity dispersion for selected stellar populations. However, their
main advantage is the ability to identify faster moving stars.

In Paper~I and here we
have used four {\it HST} imagers: WFPC2 Planetary Camera (PC1), WFPC2 Wide
Field, ACS/WFC, and WFC3/UVIS. In practice, it would be difficult to
tease apart WFPC2's PC1 data from its Wide Field contribution. Hence,
the proper-motion catalog from Paper~1 represents mixed contribution
by these WFPC2 cameras but it is possible to separate contribution
by ACS/WFC from that from WFC3/UVIS, owing to their minimal spatial overlap.
Proper
motions and associated parameters from Paper~I are marked by the
suffix ``c". There are 86,606 such entries. The new proper motions
and their parameters from ACS/WFC observations are marked by the suffix ``a" 
(210,745 entries), and those from WFC3/UVIS observations -- with ``u"
(165,737 entries). Finally, a catalog of combined proper motions
(suffix ``m") represents an attempt to calculate weighted mean proper
motions and their errors using available combinations of individual proper
motions. We did not address the cases showing clear discrepancy in proper
motion between two separate measurements, unless it is a potential runaway.
The rule of thumb in such cases is that a smaller total proper motion
is more likely to be true and the new proper motions from WFC3/UVIS
measurements are the most reliable. We note that a number of brighter stars
with the proper-motion measurement in Paper~I are missing in the new catalogs.
This is because in Paper~I we applied a universal image-cutoff threshold at the
instrumental magnitude of $m_{\rm F775W}=-14.33$, while in the study it was
$-14.0$ for long exposures by ACS/WFC and, similarly, $-14.25$ for WFC3/UVIS
frames. These changes helped to elimimate poor measurements of proper motion.

Each proper motion measurement comes with its standard error estimate, the number
of datapoints, the calculated $\chi^2$, and the goodness-of-fit probability $Q$.
Ancillary data include $VI$ photometric data from \citet{sa16},
rectangular $XY$ coordinates aligned along the RA and Dec directions,
Right Ascension and Declination (J2000). The total number of entries
in the combined and collated proper-motion catalog is 368,787.

\subsection{Potential impact of {\it Gaia} DR\,1}\label{gaia}

\begin{figure}
{\includegraphics[scale=0.75]{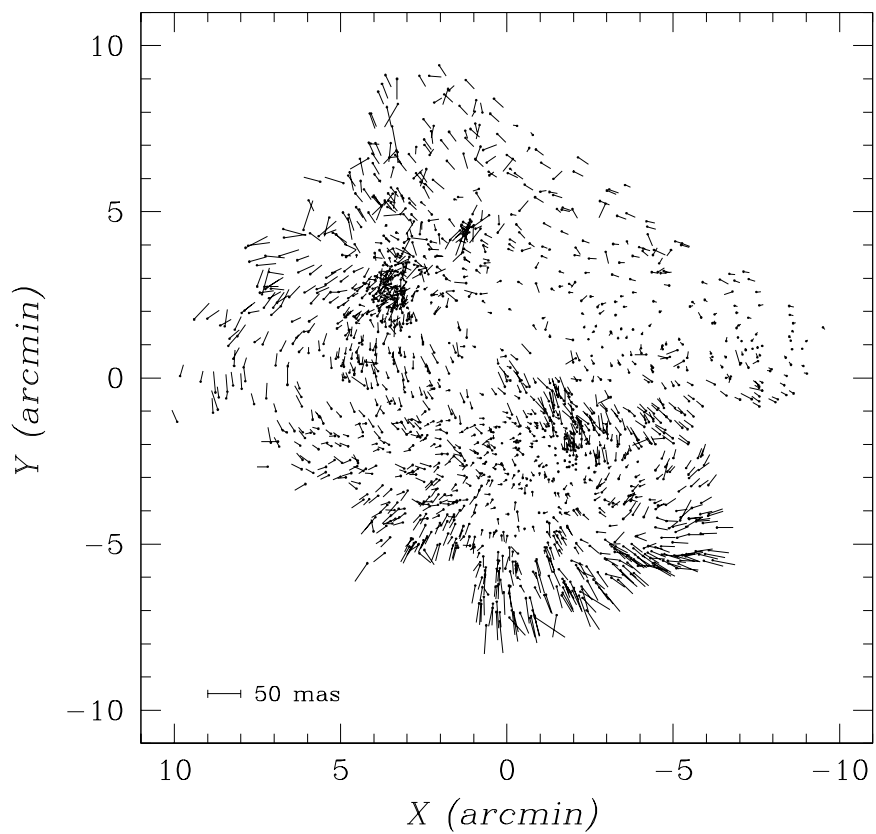}}
\caption{Positional residuals of 1738 stars common between the {\it Gaia}
DR\,1 and
our astrometric reference catalog. These residuals were obtained using a linear
least-squares adjustment. The part of the sky covered by ACS/WFC observations
is most affected by systematic errors These absolute residuals do not affect
our proper motion measurements, which are based on relative local comparisons
between observing epochs. 
}
\label{fig:gresid}
\end{figure}

The all-sky {\it Gaia} Data Release~1 \citep[DR\,1;][]{gaia} is an obvious
dataset to compare with {\it HST} measurements. We retrieved a few
thousand DR\,1 sources near the cluster R\,136. There are issues with
completeness of stars in the 30~Dor area and an apparent imbalance of
positional errors. On average, the mean positional error in Right Ascension
is $\sim$4 times smaller than that in Declination. It was decided to use
only those stars with positional errors less than 2~mas in either
coordinate, which limits the sample to $\sim$2,000 stars all
brighter than {\it G}$=$19. First, we calculated J2000 equatorial
coordinates for our astrometric reference catalog using the DR\,1
stars as a reliable representation of the International Celestial
Reference System (ICRS). We adopted a linear 6-parameter polynomial
model to transform the pixel coordinates into the tangential coordinates
of DR\,1 stars. As demonstrated by Fig.~\ref{fig:gresid}, the calculated
residuals contain significant systematics reaching up to $\sim$50~mas.
If we consider only the WFC3/UVIS positions, then the corresponding
residuals reach only $\sim$7~mas. This is a direct evidence that
the positions from observations with ACS/WFC are not free from
residual geometric distortions. A similar conclusion has been reached
from independent studies of an {\it HST} astrometric reference catalog
based on ACS/WFC observations of the globular cluster 47~Tuc
\citep{ko15}. For the sake of
proper motions in the 30~Dor area, the less than optimal accuracy of
our ACS/WFC reference catalog is not as detrimental as it may look.
The ACS/WFC CCD chip covers an area of $3\farcm4\times1\farcm7$ on the sky.
Even the long side of a chip may be shorter than the characteristic
length of correlated residuals, which is corroborated by the sub-mas
standard errors in local solutions (Table~2, Paper~I).
Since the placing of second-epoch frames relative to first epochs
normally is within a 1-2 ACS/WFC pixels, we used essentially the same
reference stars for both epochs. These mitigating factors significantly
lower the impact of an imperfect reference catalog.

We also explored whether the DR\,1 can be used as an additional epoch
to improve the precision of proper motions. Note that the timing
of our second epoch is $\sim$2014.8, while for DR\,1 it is exactly
2015.0. That near simultaneousness ensures easy comparison between two
sets of positions.  Only the WFC3/UVIS frames
and the corresponding WFC3/UVIS astrometric reference catalog were used.
It is convenient to work in the system of our astrometric reference
catalog. Therefore, we calculated the gnomonic-projection's
tangential coordinates with their zeropoint at RA\,$=84.51667$ and
Dec\,$=-69.14361$, both in decimal degrees. To emulate a WFC3/UVIS frame,
we applied the absolute scale for the F775W filter from \citet{be11}.
Small offsets and the global rotation of these
transformed {\it Gaia} coordinates
with respect to our astrometric reference catalog were eliminated by
a local linear least-squares adjustment to each DR\,1 star
using the nearest 10 common stars. This resulted in a total of 218
{\it Gaia} positions, ready to be added to the pool of WFC3/UVIS
positions. A linear unweighted least-squares fit to the available sets
of $X$ and $Y$ pixel positions as a function of time (and containing
one {\it Gaia} position) provides some insights on mixing {\it HST} and
{\it Gaia} positional data. First, there is a handful of outliers up to
$\sim$15 mas possibly indicating some structure/multiplicity in the star's image.
Second, the dispersion of {\it Gaia} positions in Declination at 
$\sim$1.6~mas is twice as large as that in Right Ascension, just
reflecting the range of listed {\it Gaia} errors for these stars.
Third, the {\it HST} WFC3/UVIS  positional uncertainty for optimally
exposed stars is nearing $\sim$0.5~mas, while the corresponding positions
from DR\,1 may have  up to four times worse positional accuracy. 
Given the large disparity of positional errors in the DR\,1 along
each of the axes, there was little to gain from incorporating them
in our dataset.

\section{DISCUSSION AND APPLICATIONS}\label{discuss}

The primary objective of this project is to provide proper motions
and the related tangential velocities
for a large sample of OB stars with measured multi-epoch LOS velocities
in the framework of teh VFTS survey. We note that, at the distance
 of the LMC, a proper
motion of 0.1~mas~yr$^{-1}$ is equivalent to 25 km~s$^{-1}$. The existing
measurements indicate the excess LOS velocities from VFTS data are in
the range of 30 to $\sim$100 km~s$^{-1}$ \citep[]{ev15,eva15,vi17}, or
$\sim$0.1 to 0.4 mas~yr$^{-1}$ in the tangential plane. 

Currently, this range appears to be well-established by the existing LOS
velocity measurements. Now we are in the position to test them with proper
motions. First, we checked for proper motions exceeding $\sim$0.4 mas~yr$^{-1}$
among all potential OB stars with emphasis on the VFTS sample. The same
exercise was done for the brighter RGB stars with $V\!-\!I>0.9$ in order to
find relatively fast moving Milky Way stars. Second, we similarly parsed
the $\sim$0.1 to 0.4 mas~yr$^{-1}$ range but now employing the calculated
and the estimated mean errors. There is a distinction between these two
types of error estimates because a small number of datapoints (n$\leq4$) in
the least-squares adjustment favors underestimated formal
proper-motion errors.
Unfortunately, many VFTS stars have this issue for our new proper motions.
We note that a fit of only three datapoints was retained for
stars with  $m_{\rm F775W}<20.0$ in order to have a proper motion estimate
near the perimeter of the 30~Dor field.

Given the fact that the accuracy of our proper motions depends on the
{\it HST} imaging instrument, we decided to divide all 481 measured VFTS
stars in three groups: proper motions from ACS/WFC (152 stars),
proper motions from WFC3/UVIS observations (183 stars), and those
from Paper~I which do not have counterparts in our new
proper motion catalog (146 stars)

\subsection{VFTS stars and Paper~I}\label{vfts_p}

In Paper~I (Table~3) we made the first attempt to identify candidate
OB runaway stars. This effort resulted in 6 such stars. Among these stars, the
largest motion along either axis is 0.40~mas~yr$^{-1}$. We note that all of
them but ID~6 have their second epochs based on ACS/WFC observations.
Only two OB runaway candidates (ID~1 and ID~6 = VFTS~285) have new proper
motions. Star~1 (\#264049) has its new proper motion incompatible with
that from Paper~I. We conclude
that it is  probably not an OB runaway. Star VFTS~285 has a consistent proper
motion from both sources, albeit it appears to have a smaller 
newly-measured motion (see Table~1).
Therefore, it remains a good candidate OB runaway.
The four remaining stars are too bright to derive credible new proper motions.
However, two of them (ID 3 and 4) have only one reliable position at each
epoch which allows us to estimate the upper limits of proper motion.
Star 3 (\#278880) has inconsistent positional data 
and in addition has a slightly asymmetric image
indicating a likely visual binary. 
Given the stated issues with our
ACS/WFC data (Sect.~\ref{promo}), it is safer to downgrade our initial
OB runaway candidates IDs 1-5 to the normal background young and massive stars.

Among the VFTS stars presented in Paper~I, there are a dozen of seemingly
fast moving stars outside the $+/-$0.4~mas~yr$^{-1}$ box centered on
zero proper motion. An examination of individual fits for the proper
motion indicates that all but one are likely spurious. Star VFTS 712,
spectral type B1\,V \citep{ev15,du15} is a nearly-equal-brightness
visual binary (\#290441 and \#290538) separated by 228~mas, and oriented
almost exactly along Declination. For both components, the proper motion
in Declination appears to be fairly large but that is not confirmed by
additional checks. Instructive is the case of VFTS 167, which appears
to have at least a 5$\sigma$ proper motion and formally would qualify for
a genuine OB runaway. However, VFTS 167 has only two good-quality datapoints
at recent 2011-2014 epochs which is not sufficient to obtain any error
estimates. Just a direct difference of available epoch coordinates indicates
a small proper motion at $\sim$0.2~mas~yr$^{-1}$ in both axes, although
with opposite sign in the RA direction to the listed one in Paper~I.
Clearly, its proper motion in Paper~I is not reliable. Such cases are
present in Paper~I for two reasons: 1) the earlier-adopted cutoff magnitude
included some slightly overexposed stellar images, and 2) archival {\it HST}
frames are placed randomly in the field of 30~Dor as opposed to
our design intended to optimize the astrometric output.

\subsection{VFTS stars and ACS/WFC measurements}\label{vfts_a}

\begin{figure}
{\includegraphics[scale=0.75]{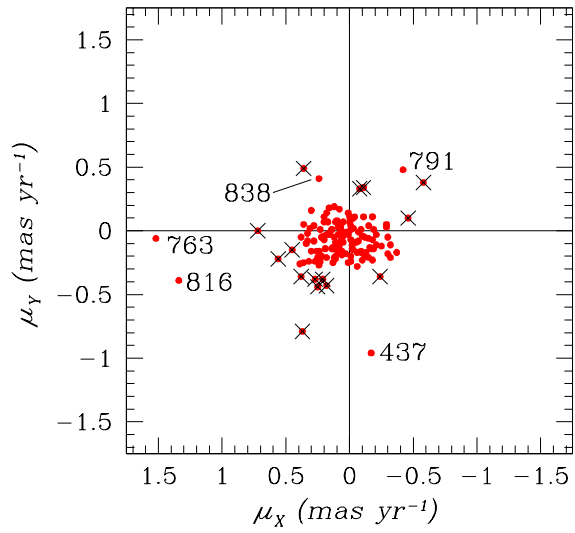}}
\caption{Vector-point diagram of ACS/WFC-based proper motions for VFTS stars.
The status of labeled stars is discussed in Sect.~\ref{vfts_a}. Inconsistent or
apparently flawed proper motions are crossed out. This diagram yields only
a single candidate OB runaway (VFTS 838).
}
\label{fig:vfts_acs}
\end{figure}

The vector-point diagram of ACS/WFC proper motions of 152 VFTS stars
(Fig.~\ref{fig:vfts_acs}) shows a fairly large scatter and a visibly
non-Gaussian distribution of the smaller motions. As stated in
Sect.~\ref{promo}, for brighter stars, new proper motions from ACS/WFC
measurements are by a factor $\sim$1.5 less precise than those obtained
from the WFC3/UVIS observations. We suspect that some lingering
systematics in the pixel positions after correcting them for geometric
distortion might be responsible for this unusual feature.
Consequently, the distribution
of proper motions within 0.4~mas~yr$^{-1}$ around zero is too broad
to allow a meaningful identification of potential relatively-slow OB
runaways. At higher proper motions, a total of five numbered stars survive
a limited scrutiny. Among them, only VFTS 838 could be a potential
OB runaway (see Table~1).
The remaining four stars are of late spectral type (G0-K7) and marginally
consistent with the LMC membership from measured LOS velocities.

\subsection{VFTS stars and WFC3/UVIS measurements}\label{vfts_u}

The sample of proper motions based upon WFC3/UVIS observations is the most
accurate and reliable source of tangential velocities available
from this project.
As such, this sample is a cornerstone in our discussion of OB runaways and
other fast moving stars. There are 183 VFTS stars in this sample. First,
we exclude VFTS 680, which is apparently a G-type Milky Way star with
large proper motion: $\mu_X=+11.15\pm0.22$ $\mu_Y=-1.17\pm0.04$ mas~yr$^{-1}$.
There is no overlap with proper motions from ACS/WFC observations,
however, there are 104 proper motion measurements common with Paper~I.
In the majority of
cases, the formal proper-motion errors are similar between Paper~I and
the WFC/UVIS sample of VFTS stars. Therefore, we can choose the {\ smallest}
proper motion as a lower limit of the expected tangential motion, if there
is a significant variation between these two sources. Normally, the
preference is given to the WFC/UVIS sample because it was
constructed to minimize systematic errors, which might be
present in the proper motions of Paper~I.

The vector-point diagram of 110 well-measured proper motions with errors
$\sigma_{\mu}\le0.13$ mas~yr$^{-1}$ is shown in Fig.~\ref{fig:vfts_uvis}.
There are two labeled VFTS stars standing apart from the general
distribution of relative proper motions around the zero motion.
Star VFTS 8 of spectral type B0.5:V(n) \citep{ev15}
appears to be a candidate runaway. VFTS 245 is a K spectral type star
\citep{ev11} and therefore is not considered. The sum of unity Gaussians
is representing the empirical 1-D proper-motion distributions in each axis
and a Gaussian fit to these distributions is provided in
Fig.~\ref{fig:vfts_uvis} (compare that to Fig.~15, Paper I). There are
small offsets, $\Delta\mu_X=-0.035$ and
$\Delta\mu_Y=+0.012$ mas~yr$^{-1}$,
which along with the distribution's Gaussian width of 0.12
mas~yr$^{-1}$ for each axis indicate that our estimates of
proper-motion errors are robust.

\begin{figure}
{\includegraphics[scale=0.75]{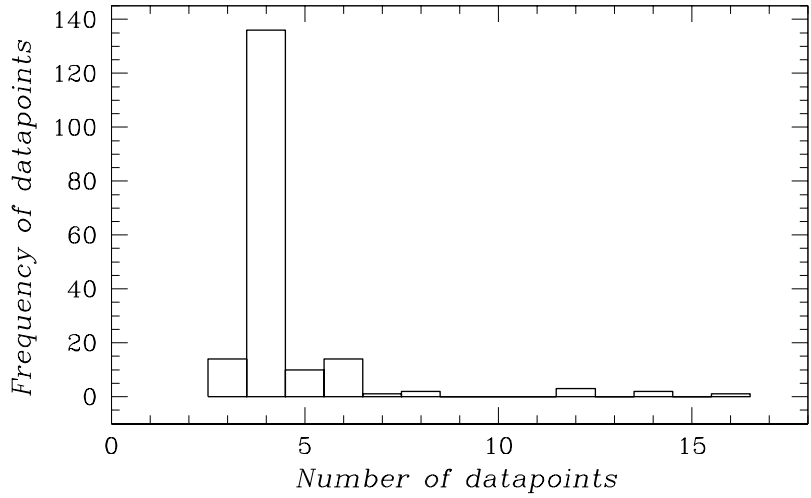}}
\caption{Histogram of the number of positional datapoints for VFTS stars
used in the WFC3/UVIS proper motion measurements.
}
\label{fig:vfts_hist}
\end{figure}

\begin{figure}
{\includegraphics[scale=0.75]{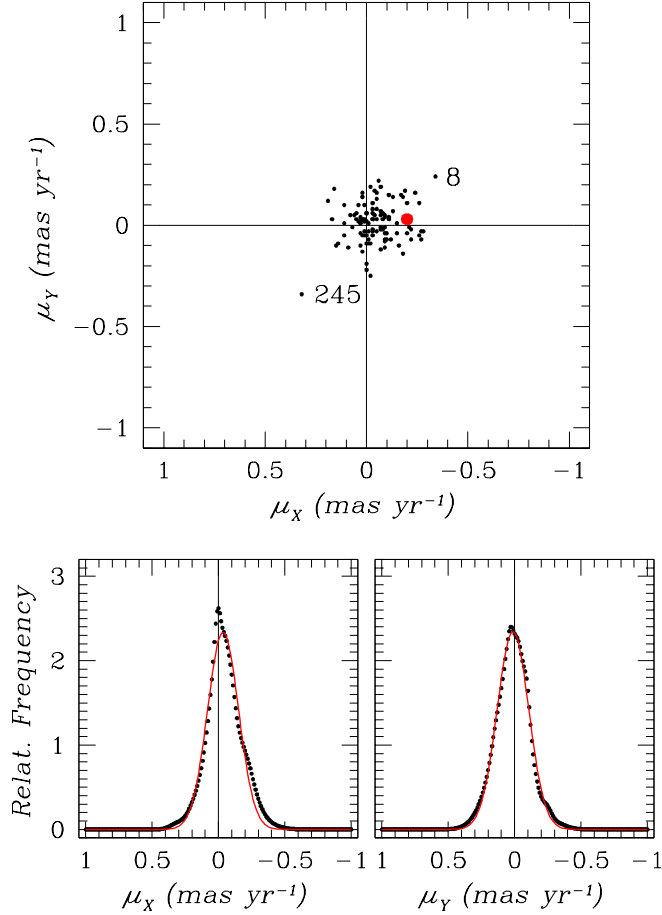}}
\caption{Proper motions of 110 VFTS stars from the WFC3/UVIS measurements. Only
proper motions with errors $\sigma_{\mu}\le0.13$ mas~yr$^{-1}$ are selected.
Upper panel: a vector-point diagram. The status of two labeled stars is
discussed in Sect.~\ref{vfts_u}. The likely OB runaway star VFTS 285 is marked
in red. Bottom panels: black points represent empirical 1-D distributions
of proper motions; red points show a Gaussian fit to these distributions
yielding a width of 0.12 mas~yr$^{-1}$ and  indicating small offsets.
}
\label{fig:vfts_uvis}
\end{figure}

In order to isolate potential OB runaways, an obvious choice seems
to be comparing each measured proper motion with its error and then
selecting 3$\sigma$ and higher significance cases. Unfortunately, the majority
of least-squares proper motion fits has a very small number
of positional datapoints (Fig.~\ref{fig:vfts_hist}), which brings on
underestimated proper-motion errors. To mitigate this deficiency,
we extracted the median error from each of the three catalogs as a function
of magnitude, in four intervals of available datapoints: $3\leq n\leq 5$,
$6\leq n\leq 7$, $8\leq n\leq 12$, and $n> 12$. 
For example, the average median error
for the VFTS stars considered here is 0.07~mas~yr$^{-1}$. 
We designed a sequence of criteria and checks that were used to identify
candidate runaway stars. It is safe
to reject the cases where the actual error of proper motion is more
than 3 times larger than its estimated median error and use the latter
as a benchmark to select potential runaway stars. In addition, we rejected
the cases when the calculated proper motion errors in each axis differ by
more than a factor of 3 and the number of available datapoints
is only 3 or 4. This eliminates the majority of unstable solutions
for proper motion.
The next step is to check the consistency of proper motion between our
catalog in Paper~I and that from WFC3/UVIS measurements. The smallest total
proper motion is thought to be more likely.
An additional criterion of reliability is the sign of the proper motion,
which should be  the same in both catalogs. The last step is a visual inspection
of the actual least-squares fit and its residuals. Finally, a runaway
candidate is visually inspected on a combined image (see Sect.~4.2 in
Paper~I).  We examined all VFTS stars which have measured proper motion
in either axis exceeding 0.17~mas~yr$^{-1}$ (equivalent to 40 km~s$^{-1}$
at the distance of LMC).
The new candidate OB runaway stars matching all our criteria,
when available, are given in Table~1. This table lists VFTS number \citep{ev11},
identifier from our electronic table,
$VI$ photometry from \citet{ci15}, spectral type, v$_{\rm LOS}$ in km~s$^{-1}$,
v$sin$i and its error in km~s$^{-1}$. Additional entries in Table~1  are
the same as in Table~3 of Paper~I.
Of these six candidate OB runaways we note that only one star, VFTS~406,
was flagged previously as a runaway from its LOS velocity (304 km~s$^{-1}$)
and high projected rotational velocity \citep{wa14},
all other candidates having LOS velocities close to that of the mean of all
OB stars in the region ($\sim$270 km~s$^{-1}$).
VFTS~406 is also noteworthy due to its high rotational velocity,
which would be consistent with the potential origin in
the binary SN scenario for runaways; that is when one component of
a binary explodes as a core-collapse
supernova and the other component attains a large kick velocity.
Furthermore, as we discuss in the next
Sect.~\ref{from_r136}, this star's direction of motion is inconsistent with an
R\,136 origin. In fact, only 4 stars in Table~1 have directions of
motion consistent with an R\,136 origin: 
VFTS~65, VFTS~219, VFTS~285 and VFTS~290, which are discussed
further below along with the non-VFTS candidate runaways from R\,136.
The remaining two runaways in this sample with directions of motion
inconsistent with an R\,136 origin, early B-type stars VFTS~8
and VFTS~838, also have high rotational velocities that might indicate
a SN ejection scenario. Another feature common to all stars listed in
Table~1 is the lack of significant variations in the measured multi-epoch
LOS velocities \citep{ev15,san13}. All of them are classed as single stars.

\subsubsection{Candidate non-VFTS young runaways from WFC3/UVIS measurements} 

The VFTS survey \citep{ev11} contains a total of 917 stars down to $V=17$.
The completeness of OB stars over the surveyed area is not well-established,
especially around the location of R\,136. Therefore, first we used $VI$ photometry
\citep{ci15} and the location of the upper main-sequence in the CMD down
to $\sim$$5 M_\sun$, equivalent to $V\sim19$, and later extended that limit to
$V\sim22$ mag. The color cut was initially adopted at $V-I<+0.75$ mag. This
selection of targets covers the majority of young O-B-A-F spectral
type main-sequence stars. Among the likely non-VFTS OB stars, only
a couple of potential runaways can be tagged but none of them convincingly.
At fainter magnitudes,
$18.0 < V < 22.0$, there are a few dozen of somewhat redder stars with
kinematics incompatible with LMC membership. To enhance the chances of
finding {\it young} main-sequence stars in the area of 30~Dor, we adopted
a tighter limit on the color cut at $V-I<+0.43$. A list of 11 potential
young lower-mass runaway stars is given in Table~2. This is the
first-ever such a list in a galaxy other than the Milky Way. However,
membership of these stars to LMC should be confirmed by other
means such as spectroscopy.

\subsection{Candidate runaway stars and the star cluster R\,136}\label{from_r136}

\begin{figure}
{\includegraphics[scale=0.80]{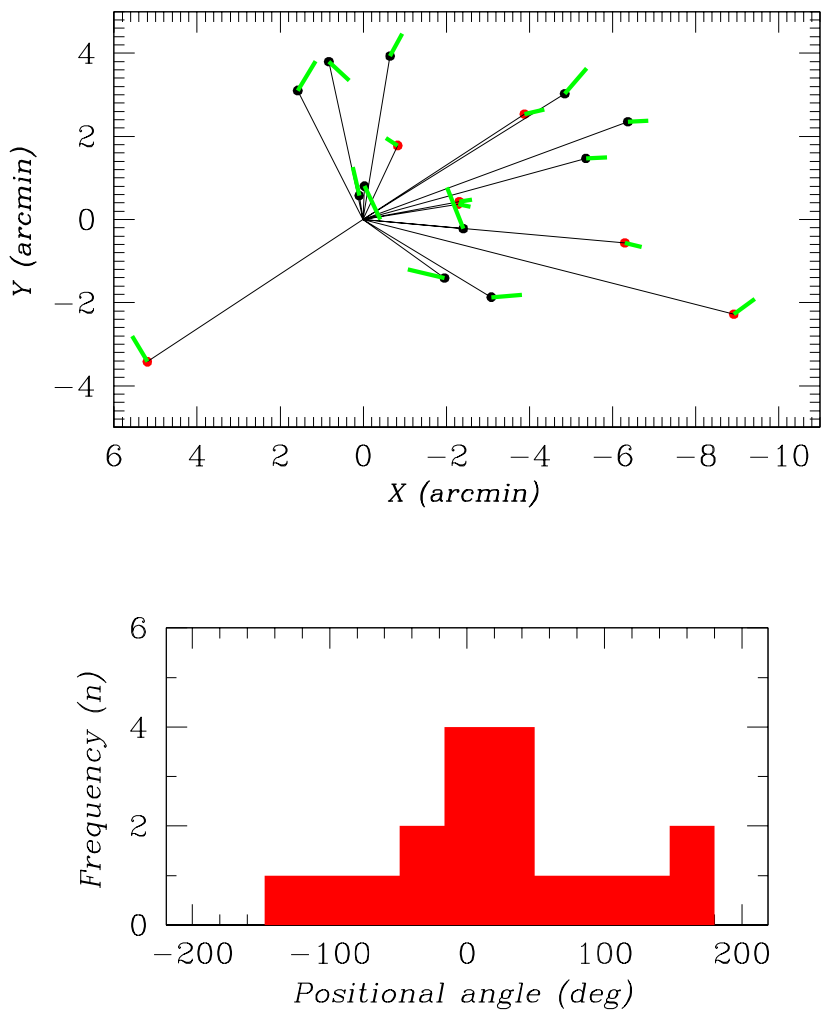}}
\caption{Distribution of 18 candidate runaway stars. Upper panel: location of
stars on the sky. Red dots indicate OB stars while black points show the likely
young A-F spectral type stars. The zeropoint of gnomonic projection is at
the center of the rich and young star cluster R\,136. Green line-segments
indicate
the direction and amount of a star's proper motion (1.5 mas~yr$^{-1}$
corresponds to $1\arcmin$). Bottom panel: histogram
of proper-motion positional angle with respect to the direction outwards
from R\,136 (thin black lines). The size of a bin is 33 deg.
There is a distinct concentration around
zero angle, which is a hallmark of genuine runaway stars.
}
\label{fig:run_polar}
\end{figure}

A principal advantage of knowing proper motions of potential runaway stars
is the ability to trace them backward to their place of origin,
if we assume that
massive star clusters are the likely nurseries of such stars. In our case,
such a place is the young and massive star cluster R\,136 which is located
at the edge of the field covered by our WFC3/UVIS observations. We explored
how the direction of proper motion vectors is oriented outwards from R\,136.
This can be characterized by the proper motion positional angle,
measured relative to the direction from R\,136. If this angle is close to
zero, then there is a high probability that a star has originated from
this cluster. Figure~\ref{fig:run_polar} show all stars from 
Tables~1,2. There is a distinct concentration
of proper-motion positional angles around zero. We used this to identify
10 plausible
escapees from R\,136 (Table~3). These stars have their
total proper motion in the range from 0.20 to 0.54 mas~yr$^{-1}$ equivalent
to 50 to 130 km~s$^{-1}$. A caveat with an actual escape velocity is that
our proper motions are relative. Due to the applied chain of transformations
leading to proper motions of stars in R\,136 and its off-the-center
position in the FOV of WFC3/UVIS, the mean proper motion of R\,136 is
biased towards zero. The formal proper motion of R\,136 is 
$\mu_X=-0.011$, $\mu_Y=+0.003$ mas~yr$^{-1}$ as calculated using the brighter
and bluer stars ($V<22$, $V-I<0.5$) within one arc-minute around the
center of R\,136.
Inspecting the rms distribution of specific WFC3/UVIS frame transformations
containing the cluster (see Fig.~\ref{fig:rms_all}), the estimated
uncertainty in the proper motion of R\,136 appears to be less than
$\sim$0.05 mas~yr$^{-1}$.

The angular distance of an escapee and its proper motion provide the time
when the star left R\,136. Thus, the slowest escapees -- VFTS~285 and
VFTS~290 -- have moved to the current position in 0.67~Myr. Star VFTS~65 has
traveled 1.37~Myr and has covered 92~pc away from R\,136. If the isochrone
age of R\,136 age is considered \citep[3~Myr,][]{ci15}, then a potential
early escapee with a mass of $\sim$$20 M_\sun$ or higher may have
traveled as far as $\sim$200~pc away from R\,136, equivalent to the length of
$\sim14\arcmin$ in the tangential plane. We note that the LOS velocities of
VFTS stars imply the runaway upper limit at $\sim$100 km~s$^{-1}$ \citep{vi17}.
That would result in a distance of $\sim$300~pc ($\sim20\arcmin$)
over 3~Myr.  {\it Gaia} proper motions will enable searches for ejected
runaways from R\,136 at these greater distances from the cluster.

For those stars with position angles consistent with ejection from R\,136 
that are also in the VFTS sample (stars VFTS~65, 219, 285 and 290),
we can compare their evolutionary ages \citep{schneider18}
with the age of the central cluster, estimated to be 1.5 Myr
\citep{crowther16}.
The O-type stars VFTS~65 and 285 have ages estimated at 2.4 and 1.9 Myr,
respectively with uncertainties of approximately 1 and 2 Myr,
hence they have ages consistent with that of R\,136. The remaining
confirmed O-type star, VFTS~290, has an age of 4.7 Myr with
an uncertainty of approximately 0.5 Myr. Furthermore,
from the analysis of \citet{sa17} it appears that to match the age R\,136 
this object would have to be a $\sim$3$\sigma$ outlier.  
Since this star has an age that is more consistent with the surrounding 
NGC~2070 cluster it may well that this runaway is the product of the binary
supernova mechanism \citep{bl61} and originated in that region. 
The case of VFTS~219 is even more problematic as this mid-B star has
an estimated age of 77 Myr and, while the uncertainty is $\sim10$ Myr,
this star is inconsistent with previous membership of R\,136.
We suggest that VFTS~219 (and potentially even VFTS~290) might be
an interloper in the sense that it is a
field runaway star that by chance has a position angle consistent
with ejection from R\,136.  Indeed, referring to the histogram
 in Fig.~\ref{fig:run_polar} one
can see that we expect approximately one star per 33 degree bin of
position angle, irrespective of the bin's orientation with respect to R\,136. 
In the central 3 bins we therefore expect about 3 stars
that might be interlopers in our sample.

Among the more massive escapees, we note a pair of main sequence
O-stars (VFTS 285 \& VFTS 290) which are separated only by $4\farcs27$.
Their total proper motion is identical but not the LOS velocities and
projected rotation velocities: 228 and 269 km~s$^{-1}$ \citep{san13};
$v\sin\,i$: 600  and 40 km~s$^{-1}$ \citep{sa17}. According to \citet{ev15},
the mean LOS velocity of NGC~2070, which also contains R\,136,
is 271 km~s$^{-1}$. From the perspective of LOS velocities, only VFTS 285
appears to be a runaway star, albeit of a slow variety.
However, our proper motion measurement of these stars
appears to be fairly robust, hence confirming their runaway status.
It is unlikely that this pair is a wide physical binary due to the very
large current  projected separation in the tangential plane at $\sim$213,000 AU.

Another potential runaway pair, \#371614 and VFTS~662, is even less understood.
In Table~2, star \#371614 is listed as a candidate runaway.
Its orientation angle with respect to R\,136 is $-144$ deg. Formally, that
large angle rules out any connection to this star cluster. However, the
derived proper motion appears to be biased by the nearby and more luminous
VFTS 662 (separated by  $\sim$$0\farcs3$), which is missing in our catalogs.
According to \citet{ev15}, VFTS 662 has its LOS velocity at 251 km~s$^{-1}$.
It is expected that {\it Gaia} DR\,2 will help to clarity the status of
this pair. 

\begin{figure}
{\includegraphics[scale=0.75]{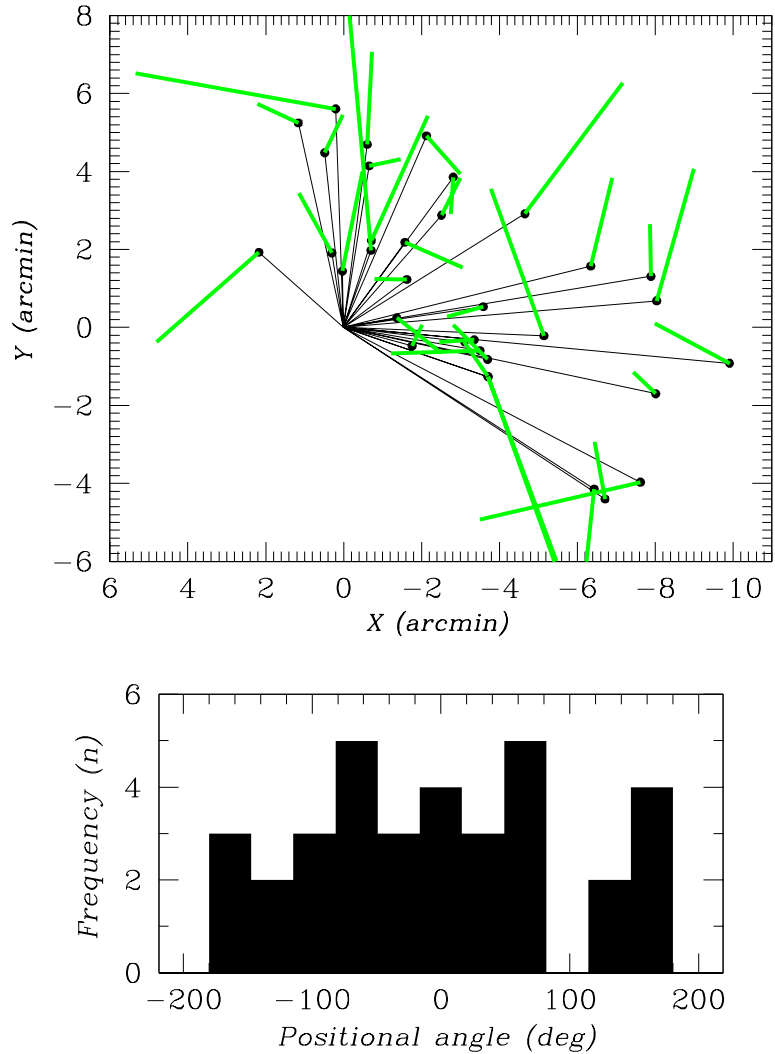}}
\caption{Distribution of fast-moving field stars. This plot is identical
to Fig.~\ref{fig:run_polar} but addresses the redder main-sequence stars
which on average have much higher proper motions and a random distribution of
positional angles. The proper-motion amplication factor used for
visualization is identical to that used in Fig.~\ref{fig:run_polar}. 
}
\label{fig:field_polar}
\end{figure}

An additional argument that our candidate escapees from R\,136 are real comes
from the distribution of redder main-sequence ($0.43<V-I<0.75$) runaway
stars down to $V=22$ mag. There are 34 such stars shown in
Fig.~\ref{fig:field_polar}. The histogram of positional angles implies a lack
of concentration at the zero orientation angle. We note that for the faster-moving
stars this concentration would be much sharper due to the greatly diminished
impact of proper motion errors. In addition, the amplitude of the total proper
motion in this sample is much larger -- from 0.4 to more than 4 mas~yr$^{-1}$,
indicating very different kinematics.

\begin{figure}
{\includegraphics[scale=0.75]{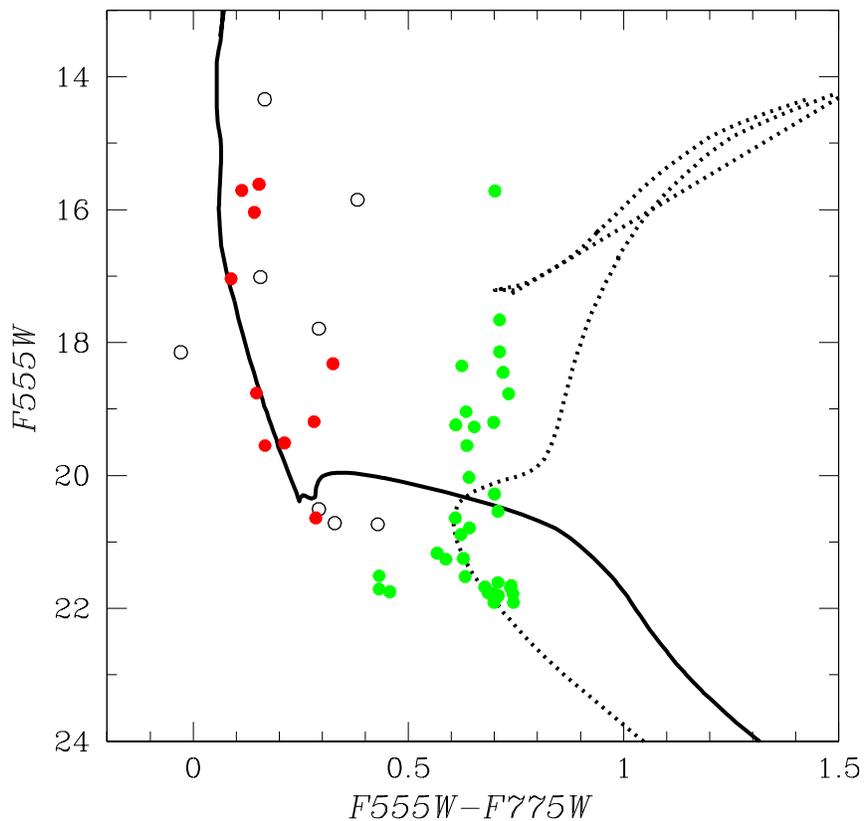}}
\caption{Color-magnitude diagram of the stars plotted in
Figs.~\ref{fig:run_polar},\ref{fig:field_polar}. Red points -- all stars from
Table~3; open circles -- candidate O-B-A-F runaways not
kinematically-associated with R\,136; green points - fast moving field stars.
Solid curved line is the 3~Myr isochrone for the cluster R\,136,
adopting $E(B-V)=0.3$. The dotted line shows an isochrone with the following
parameters: 12~Gyr old, metal-poor [Fe/H]$=-1.5$ dex, $E(B-V)=0.05$,
and placed at 20 kpc from the Sun.
}
\label{fig:cmd_run}
\end{figure}

These two sets of kinematially-selected stars also have a very different
placement on the $V-I$ vs. $V$ CMD (Fig.~\ref{fig:cmd_run}) where the
interchangeable axes are labeled {\it F555W} vs. {\it F555W-F775W} in
accordance with \citet{ci15}. The 10 escapees are well-separated from
fast-moving field stars. The stars likely escaped from R\,136 are located
along the main sequence of this young cluster down to its pre-main-sequence.
As expected, the remaining 8 candidate O-B-A-F runaways not emanating
from R\,136 are fairly close to the location of R\,136 escapees.
The fast-moving field stars form a vertical
sequence, similar to that toward the north Galactic pole \citep{re93}.
Apparently, these are Milky Way halo stars at various distances from
the Sun. To illustrate this point, we overplotted an appropriate
isochrone which fits well the most distant stars, and which are also
kinematically slowest in our sample as indicated by their total proper
motion. There is
an ambiguity, though, to which population we should assign three
isolated stars at $V\sim21.7$ and $V-I\sim0.43$. Therefore, their
status remains undefined.

\begin{figure}
{\includegraphics[scale=0.75]{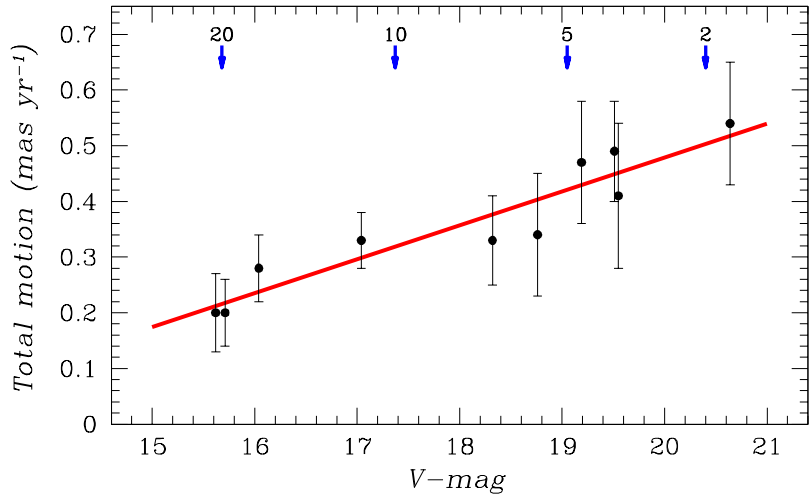}}
\caption{Total motion of runaway stars consistent with escaping from R\,136. There is
a correlation of total motion with the stellar magnitude which acts
as a proxy of stellar mass. The latter in solar units is indicated
by a down-pointed arrow at four locations of the magnitude axis.
The slope of this distribution is $0.061\pm0.008$ mas~yr$^{-1}$ per unit
magnitude.
} 
\label{fig:escape}
\end{figure}

There is an intriguing relationship for the likely-escaped stars from
R\,136 -- their total proper motion (see Table~3) is
correlated with the apparent $V$-magnitude (Fig.~\ref{fig:escape}). Since these
stars are located approximately at the same distance from the Sun, their
apparent magnitudes become akin to absolute magnitudes and can be used
as a proxy of mass for main-sequence stars.
Clearly, less massive stars are moving faster, which is contrary to what
is expected in most dynamical ejection scenarios
\citep[e.g.,][]{oh15,oh16} where the more massive runaways tend to
have higher velocities. Perusing various simulations
performed by \cite{oh16}, it appears that certain models do predict a trend
of increasing velocity with decreasing mass.  For example, in those
scenarios that assume a random pairing of binary masses among the initial
conditions that trend is further strengthened when the initial conditions
assume the cluster is not mass-segregated (models MS3RP and NMS3RP of
 \cite{oh16}).
However, we also note that the absence of faint runaways with smaller
proper motions in Fig.~\ref{fig:escape} is somewhat baffling.
For stars fainter than $V=19$, the detection limit in total proper motion
is $\sim$0.3 mas~yr$^{-1}$. The range 0.2-0.3 mas~yr$^{-1}$ for these stars
was not carefully explored owing to the large number of impostors and our
inability to identify bona fide cases. Therefore, it is appropriate to assert
that we did not find any faster-moving massive escapee, which could
match the maximum escape velocities of less massive stars. 

A potential issue with the interpretation of this correlation with the mass
might be neglecting the contribution of excess LOS velocity relative
to the mean motion of R\,136. In fact, for three massive stars out of four,
that contribution is negligible (see Table~1). For star VFTS 285, the total
escape velocity would be an equivalent of 0.27 mas~yr$^{-1}$. That alone cannot
change significantly the slope shown in Fig.~\ref{fig:escape}.
VFTS 285 has an extremely high v$sin$i estimate which makes the measurement
of LOS velocity nontrivial. However, \citet{san13} provide 7 independent
and mutually consistent estimates of LOS  velocities for VFTS 285.

The same trend as a function of $V$-magnitude can be detected among
the 8 candidate
young O-B-A-F runaways not related to R\,136. However, with the exception of
VFTS~406, their total proper motion is significantly higher
(by 0.1-0.15 mas~yr$^{-1}$, or 25-40 km~s$^{-1}$) than that shown in Fig.~\ref{fig:escape}.
This is puzzling considering that the proper-motion errors in both samples
are similar. 
It is tempting to identify this sample of runaways and, as discussed above,
even some stars in the R\,136 runaway sample as the products
of the binary supernova ejection (BSE) scenario. These would be relatively fast runways,
with tangential speeds in the range 50-100 km~s$^{-1}$. It is not straightforward
to estimate the fraction of OB stars that are BSE runaways given the 
incompleteness of detections in the ACS/WFC data,  the bright magnitude
limit that excludes many O-stars, and the unknown incompleteness of the
astrometric catalog. However 4 of our 8 BSE candidates are OB-type
stars, and we detected 481 known OB stars in our catalog. This implies an
upper limit of around 1\% as the fraction of fast OB runaways. Given the 
various caveats referred to above this is not inconsistent with 
model predictions of runway fractions of at most a few percent 
\citep{eldridge11,renzo18} although their definition of a runaway adopts
a peculiar velocity of  greater that 30 km~s$^{-1}$.

However we refrain here from deeper analysis and interpretation for two
reasons: 1) our best proper motions cover a limited
$\sim15\arcmin\times7\arcmin$ area only partially overlapping with R\,136;
and 2) our observations cannot provide reliable proper motions for stars
brighter than $V\sim15$, while luminous O-type supergiants in the 30 Dor area
can be as bright as magnitude $V\sim11$ \citep{se99}. A full analysis and
interpretation is planned once the sample can be augmented by new {\it Gaia}
DR2 proper motion measurements.

\section{CONCLUSIONS}

We derived a new proper motion catalog based on dedicated WFC3/UVIS and
parallel ACS/WFC observations. Combined with the data from our pilot archival
study, we provide proper motions for 368,787 stars in the region of 30~Dor.
Among the stars brighter than $V\sim22$ mag, a number of fast moving stars
are identified. A total of 10 runaway stars  have proper 
motion directions consistent with an origin
in the young and massive star cluster R\,136, albeit roughly one third 
of these may be chance alignments. 
Our WFC3/UVIS observations
provide the necessary accuracy to detect reliable proper motions in the range
of $\sim$0.15 to 0.5 mas~yr$^{-1}$ where the O-B-A-F runaways were identified. 
This is a unique sample of runaway stars allowing us to better understand
the process of dynamical ejection of stars in very young clusters, and to put
realistic limits on the rate of such ejections.

In summary, the main achievements of this study are:

\begin{enumerate}
\item We detail an empirical approach to account for differential
Charge-Transfer Inefficiency, which is similar for both {\it HST}
imaging instruments -- ACS/WFC and WFC3/UVIS.

\item We calculated relative proper motions for a total of 368,787 stars
down to $V\sim 25$. This second high-accuracy proper motion catalog
in the region of 30~Dor is instrumental to explore young fast-moving stars.

\item We did not find any fast moving OB star
(with total proper motion $\mu>$0.4 mas~yr$^{-1}$,
equivalent to $\sim$100 km~s$^{-1}$)
among a total of 481 measured VFTS stars. That may rule out some suggested
pathway(s) for creating runway stars in clusters. We predict that all the more
massive runaways
(mass higher than  $\sim$$20 M_\sun$), originating from the young and
massive star cluster R\,136, will be found within 200-300~pc from R\,136.

\item All candidate OB proper-motion runaway stars are single and most
of them have high projected rotation velocities, including a record-high
VFTS 285 at v$\sin$i$=600$ km~s$^{-1}$.

\item There is convincing evidence that a number of O spectral-type down
to F-stars have escaped from R\,136. It appears that these escapees have
tangential velocities correlated with apparent magnitude, which points to
the stellar mass as a driver for the  escape velocity distribution. 

\end{enumerate}

\acknowledgements  The authors gratefully acknowledge grant support
for programs GO-12499, GO-12915, and GO-13359,
provided by NASA through grants from the Space
Telescope Science Institute, which is operated  by the Association of
Universities for Research in Astronomy, Inc., under NASA contract
NAS~5-26555. We thank co-investigators Nate Bastian and Eli Bressert
for their support in the construction of this proposal.
IP thanks Jay Anderson, Terrence Girard, and Rosemary Wyse
for insights  on astrometry from saturated images, discussions
on error propagation issues, and for clarification on the properties of
various stellar populations. DJL would like to acknowledge the critical and
enthusiastic contribution made by our departed and greatly missed colleague,
Nolan Walborn.

{\it Facilities:} \facility{Hubble Space Telescope}

\newpage

\newpage

\newpage

\newpage

\newpage

\newpage

\newpage

\newpage

\newpage

\newpage

\newpage

\newpage

\clearpage
\begin{sidewaystable}
\scriptsize{
\centering
\begin{tabular}{cccclccccccccccc}
\multicolumn{16}{c}{\textsc{\bf Table 1.} Candidate OB proper-motion runaway stars}
\\
\hline\hline
VFTS & Ident & $V$ & $V-I$ & Sp. Type & v$_{\rm LOS}$ & v$sin$i  &  $\mu_X$ & $\mu_Y$ & $\sigma_{\mu_{X}}$ & $\sigma_{\mu_{Y}}$ & $\chi^2_{X}$ & $\chi^2_{Y}$ & $Q_X$ & $Q_Y$ & N$_{\rm frame}$\\
\hline
8 & 188822 &   17.014 & 0.156 & B0.5$\,$V(n) & 271 & 241$\pm$15 &  $-$0.34 & \phs0.24 & 0.10 & 0.10 & 0.27 & 0.28 & 0.899 & 0.899 & 6 \\
65 & 259107 &  16.036 & 0.142 & O8$\,$V(n) & 268 & 162$\pm$20 &   $-$0.27 & $-$0.07 & 0.06 & 0.09 & 0.06 & 0.14 & 0.814 & 0.706 & 3 \\
219 & 349614 & 17.043 & 0.088 & B3-5$\,$III-V & 282 & 220$\pm$19 &  $-$0.32 & \phs0.07 & 0.04 & 0.15 & 0.04 & 0.62 & 0.996 & 0.646 & 6 \\
285 & 294979 & 15.616 & 0.052 & O7.5$\,$Vnnn & 228 & 600$\pm$?? & $-$0.20   & \phs0.03 & 0.07 & 0.12 & 0.13 & 0.37 & 0.877 & 0.687 & 4 \\
290 & 292849 & 15.708 & 0.113 & O9.5$\,$IV & 269 & $<$40 &   $-$0.20 & $-$0.04 & 0.06 & 0.06 & 0.09 & 0.09 & 0.914 & 0.915 & 4 \\
406 & 332741 & 14.340 & 0.166 & O6$\,$Vnn & 304 & 356$\pm$30 &   \phs0.19 & \phs0.12 & 0.05 & 0.06 & 0.06 & 0.06 & 0.946 & 0.944 & 4 \\
838 & 143204  & 15.852 & 0.382 & B1:$\,$II(n) & 263 & 239$\pm$23 &   \phs0.24 & \phs0.41 & 0.20 & 0.11 &1.03 & 0.29 & 0.355 & 0.750 & 4 \\
\hline
\end{tabular}}
\end{sidewaystable}

\begin{sidewaystable}
\scriptsize{
\centering
\begin{tabular}{ccccccccccccccc}
\multicolumn{12}{c}{\textsc{Table 2. Candidate young A-F proper-motion runaways\label{tab:afrun}}}\\
\hline\hline
Ident & $V$ & $V-I$ & $RA$ (deg) &
$Dec$ (deg) & $\mu_X$ & $\mu_Y$ & $\sigma_{\mu_{X}}$ & $\sigma_{\mu_{Y}}$ & $\chi^2_{X}$ & $\chi^2_{Y}$ & $Q_X$ & $Q_Y$ & N$_{\rm frame}$\\
\hline
204988 & 19.509 & \phs0.212 & 84.5362650 & $-$69.1332743 & $-$0.49 & \phs0.04 & 0.09 & 0.12 &0.25 & 0.50 & 0.958 & 0.810 & 8 \\
223800 & 17.790 & \phs0.292 & 84.5887628 & $-$69.1256231 & \phs0.59 & \phs0.14 & 0.04 & 0.08 &0.03 & 0.12 & 0.971 & 0.888 & 4 \\
271782 & 20.508 & \phs0.292 & 84.5679256 & $-$69.1058476 & \phs0.26 & \phs0.66 & 0.10 & 0.07 &0.72 & 0.34 & 0.721 & 0.978 & 13 \\
299597 & 19.188 & \phs0.281 & 84.6849573 & $-$69.0926151 & \phs0.10 & \phs0.46 & 0.09 & 0.11 &0.43 & 0.64 & 0.901 & 0.747 & 10 \\
306820 & 20.717 & \phs0.329 & 84.6791949 & $-$69.0887983 & $-$0.26 & $-$0.54 & 0.12 & 0.06 &0.69 & 0.14 & 0.739 & 0.999 & 12 \\
325244 & 18.757 & \phs0.147 & 84.4301820 & $-$69.0775362 & $-$0.34 &\phs0.02 & 0.11 & 0.06 &1.22 & 0.41 & 0.248 & 0.978 & 17 \\
346142 & 18.324 & \phs0.325 & 84.3832657 & $-$69.0627707 & $-$0.33 &\phs0.02 & 0.08 & 0.06 &0.47 & 0.34 & 0.936 & 0.981 & 12 \\
358858 & 20.637 & \phs0.285 & 84.4541695 & $-$69.0516393 & $-$0.35 &\phs0.41 & 0.11 & 0.11 &0.77 & 0.75 & 0.778 & 0.704 & 14 \\
359942 & 20.735 & \phs0.429 & 84.7538538 & $-$69.0505241 & $-$0.29 &\phs0.47 & 0.17 & 0.10 &1.31 & 0.43 & 0.234 & 0.904 & 14 \\
371614 & 18.147 & $-$0.029 & 84.7191400 & $-$69.0389415 & $-$0.33 & $-$0.30 & 0.07 & 0.10 &0.27 & 0.53 & 0.977 & 0.833 & 10 \\
373715 & 19.546 & \phs0.167 & 84.6501694 & $-$69.0366983 & $-$0.19 &\phs0.36 & 0.11 & 0.14 &0.90 & 1.35 & 0.545 & 0.183 & 14 \\
\hline
\end{tabular}}
\end{sidewaystable}

\clearpage
\begin{deluxetable}{ccccc}
 \tablecolumns{5}
 \tablenum{3}
 \tablewidth{0pt}
 \tablecaption{Likely-escaped stars from R\,136}
 \tablehead{
    \colhead{Ident}            &
    \colhead{X (arcmin)}    &
    \colhead{Y (arcmin)}    &
    \colhead{Angle (deg)} &
    \colhead{Total pm (mas~yr$^{-1}$)}
}
\startdata
204988 & $-$3.0804 & $-$1.8681  & \phs35.9 & 0.49 \\
259107 & $-$6.3002 & $-$0.5612 &    $-$9.4 & 0.28 \\
292849 & $-$2.2820 & \phs0.3702 & $-$20.5 & 0.20\\
294979 & $-$2.3160 & \phs0.4293 & $-$2.0 & 0.20 \\
299597 & \phs0.0976 & \phs0.5751 & \phs2.6 & 0.47 \\
325244 & $-$5.3612 & \phs1.4689 & $-$12.0 & 0.34 \\
346142 & $-$6.3708 & \phs2.3503 & $-$16.8 & 0.33 \\
349614 & $-$3.8849 & \phs2.5365 & $-$20.8 & 0.33 \\
358858 & $-$4.8530 & \phs3.0247 & \phs17.6 & 0.54 \\
373715 & $-$0.6489 & \phs3.9299  & $-$18.4 & 0.41 \\
\enddata
\end{deluxetable}

\end{document}